\documentclass[submission,copyright,creativecommons,sharealike]{eptcs}
\providecommand{\event}{ICLP 2019} 
\usepackage{underscore}           

\usepackage{pgfplots}
\usepgfplotslibrary{statistics}
\usepackage{mathptmx}
\usepackage{subcaption}
\usepackage{todonotes}
\usepackage{csquotes}

\usepackage{url}
\usepackage[capitalize]{cleveref}

\usepackage{fancyvrb}
\usepackage{xspace}

\usepackage{tikz}
\usepackage{pgfplots}
\usetikzlibrary{patterns}
\usetikzlibrary{fit,shapes,arrows,arrows.meta,decorations.pathreplacing}

\usepackage{booktabs}

\newcommand{\cov}[1]{\textsc{Cov}~#1}
\newcommand{\ccov}[1]{(\cov{#1})}
\newcommand{\swipl}{SWI-Prolog\xspace}
\newcommand{\exm}[1]{{\scriptsize \texttt{#1}}}

\title{Prolog Coding Guidelines: Status and Tool Support}
\def\titlerunning{Prolog Coding Guidelines: Status and Tool Support}

\author{
Falco Nogatz
\institute{University of W\"{u}rzburg, Germany}
\email{falco.nogatz@uni-wuerzburg.de}
\and 
Philipp K\"{o}rner
\institute{University of D\"{u}sseldorf, Germany}
\email{p.koerner@hhu.de}
\and
Sebastian Krings
\institute{Niederrhein University of Applied Sciences, Germany}
\email{sebastian.krings@hs-niederrhein.de}
}
\def\authorrunning{F. Nogatz, P. K\"{o}rner \& S. Krings}

\begin{document}

\maketitle

\begin{abstract}
The importance of coding guidelines is generally accepted throughout developers of every programming language.
Naturally, Prolog makes no exception.
However, establishing coding guidelines is fraught with obstacles:
Finding common ground on kind and selection of rules is matter of debate;
once found, adhering to or enforcing rules is complicated as well, 
not least because of Prolog's flexible syntax without keywords.
In this paper, we evaluate the status of coding guidelines in the 
Prolog community and
discuss to what extent they can be automatically verified.
We implemented a linter for Prolog and applied it to several packages 
to get a hold on the current state of the community.
\end{abstract}


\section{Introduction}
The importance of coding guidelines is generally accepted throughout developers of every programming language.
Naturally, Prolog makes no exception.
Hence, different attempts at establishing coding guidelines for Prolog have been made.
The most recent work towards unified coding guidelines for Prolog has 
been made by
Covington et al.~(2012)
as well as
McLaughlin~(1990).
Further comments and additions to McLaughlin's style guide have been 
posed by
Kaplan~(1991),
who focused on 
the presentation and implementability of the given rules.

However, establishing coding guidelines is fraught with obstacles:
Finding common ground on kind and selection of rules is matter of debate;
once found, adhering to or enforcing rules is complicated.
Furthermore, not all rules deemed sensible can be automatically verified.

Covington et al. suggested the usage of automated tools to 
improve readability~\cite[Sec.~2.17]{dblpjournals/tplp/CovingtonBOWP12}.
While the authors suggested numerous tools able to reformat source code, we want to go move beyond reformatting 
and automatically verify a greater extent of the coding guidelines 
suggested.
In order to provide such a tool, we evaluate the status of coding guidelines in the Prolog community,
discuss to what extent they can be automatically verified
and suggest possible ways to move forward.
We try to assert the impact of existing guidelines empirically and enquire if there is consensus on how to 
format Prolog code.

Following, in \cref{sec:automation}, we discuss how well 
the rules introduced by Covington et al. can be checked automatically.
In particular, we define a subset
we deem suitable for 
automatic verification and discuss why others cannot be checked by a 
software tool.
For the ones where automation is feasible, we implemented a style 
checker and linter which we introduce in \cref{sec:linter}.
The linter is applied to a selection of publicly available Prolog 
files from multiple sources.
In \cref{sec:report}, we present the current state of coding guidelines 
and to what extent the community adheres to them.
Finally, in Sections~\ref{sec:related-work} and~\ref{sec:conclusion} 
we discuss related and future work and conclude.

\section{Applicability of Automatic Checking to Coding Guidelines}%
\label{sec:automation}

The coding guidelines given by Covington et al.~(2012) and 
McLaughlin (1990)
involve different aspects of a Prolog program, ranging from formatting 
instructions to naming of variables and predicates as well as advice on 
implementation aspects such as the usage of cuts.
In the following, we discuss some rules stated 
in~\cite{dblpjournals/tplp/CovingtonBOWP12}.
For the sake of simplicity, we refer to them as \textsc{Cov}, with
\cov{2.1} representing the coding guideline of Section~2.1.
\Cref{tbl:implementationstatus} lists which rules can be automated and to which extend they are already implemented in our linter.
To give an impression of why rules can or cannot be automated, we discuss some of them below.

\begin{table}[t]
  \caption{Implementation Status of \cov Rules}
  \label{tbl:implementationstatus}
  \begin{tabular}{llll}
    \toprule
    Category & Implementable & Not-Implementable & Implemented \\ \midrule
    Layout & 2.1 -- 2.9, 2.14 -- 2.17 & 2.10 -- 2.13 & 2.1 -- 2.7 \\
    Naming conventions & 3.1, 3.4, 3.10, 3.12 & 3.2, 3.3, 3.5 -- 3.9, 3.11, 3.13, 3.14 & 3.1, 3.4 \\
    Documentation & 4.1 & 4.2 -- 4.5 & - \\
    Language idioms & 5.3 & 5.1, 5.2, 5.4 -- 5.13 & -\\
    Development, debugging,\\ and testing & 6.5, 6.7 & 6.1 -- 6.4, 6.8 -- 6.18 & - \\
    \bottomrule
  \end{tabular}

\end{table}

Certain rules can easily be implemented, in particularly those involving source code 
layout.
\cov mentions indentation with spaces instead of tabs, 
consistent indentation, limits regarding the length of lines or the 
number of lines in a predicate as well as the use of spaces and 
newlines.
In addition, \cov advises a decision on consistent ways to write 
if-then-else and disjunction.
However, even certain layout rules cannot easily be verified 
automatically since they rely on personal opinion or subjective 
assessment, for instance:
\begin{itemize}
    \item[--] Use layout to make comments more readable~\ccov{2.10}.
    \item[--] Avoid comments to the right of a line of code, unless 
    they are inseparable from the lines on which they 
    appear~\ccov{2.11}.
\end{itemize}
We argue that both rules are unsuitable for automatic linting.
Readability is a highly personal evaluation, depending on programmer's 
taste and not evaluable without understanding a comment's content.
While language processing techniques might be able to capture parts of 
intention and mood using, for instance sentiment 
analysis~\cite{MEDHAT20141093}, we see no appropriate technique for 
incorporation in a linting tool.

Similar problems occur when trying to verify naming 
conventions~\ccov{3}.
While we can of course check the consistent use of underscore-style or camel case as well as the consistent naming of identifiers, we again face rules which are not easily implementable:
\begin{itemize}
    \item[--] Rules regarding the pronunciation of 
    identifiers~\ccov{3.2, 3.3}.
    \item[--] Rules taking into account the semantics of a predicate, 
    such as assigning certain names to predicates representing 
    relations vs. 
    those to be understood procedurally~\ccov{3.6, 3.7},
    \item[--] Add the types on which a predicate operates to the 
    predicate name~\ccov{3.13}.
\end{itemize}
While again language processing techniques might be able to solve the first problem, capturing the intended semantics of a predicate is complicated at best.
Since Prolog is an untyped language, the types on which a predicate 
operates cannot be computed in general.
Ciao Prolog includes an assertion language that allows annotating Prolog programs with modes, types and further information that could be used to improve coverability of rules~\cite{Puebla2000,Puebla98aframework}.
In other cases, an external library such as 
\textit{plspec}~\cite{plspec} for specifying types or other constraints 
on variables could be integrated into a style checker.

Rules regarding availability of documentation~\ccov{4} can only be verified with respect to the existence of documentation and 
comments.
However, enforcing the existence of comments might just drive programmers to add empty or meaningless comments.

The guidelines in \cov{5} discuss language idioms and highlight the 
importance of certain ways to write predicates, going into detail on 
techniques such as tail recursion.
While these usually involve common patterns that can be checked 
syntactically, deciding on when to use them remains impossible for a style checker.
This holds true for all rules regarding language idioms:
\begin{itemize}
    \item[--] Deciding whether a predicate is steadfast is impossible if we aim
    for absolutely correct classification. However, a common pattern that often 
    collides with steadfastness is the unification of a parameter after a cut. In that case, it is impossible to 
    backtrack into another rule that might be more appropriate for the 
    given parameter.
    \item[--] Without further annotations regarding type or mode of 
    inputs, it is impossible to decide whether input arguments have 
    been placed before output arguments. Again, we refer to the 
    different articles regarding type systems and mode checkers for 
    Prolog~\cite{mycroft1984polymorphic,mercurytypes,plspec}.
    \item[--] Deciding whether a cut is red or green is impossible in general. Hence, a linter cannot 
    use different rules for red and green cuts as suggested 
    by \cov{5.4}.
    \item[--] \cov{5.7 and 5.8} suggest avoiding \texttt{append/3} as 
    much as possible for performance reasons.
    However, we do not think its sensible to mark its usage
    as stylistic errors in general.
    The same holds true for \texttt{asserta/1, assertz/1} and \texttt{retract/1}
    as \cov{5.10} discusses.
    \item[--] Of course a linter can mark occurrences of body recursion 
    and suggest to use tail recursion \ccov{5.9}.
    At the same time, we cannot estimate whether a reimplementation will be worthwhile.
    \item[--] Operating in badges (\cov{5.11}) and using built-in sorting algorithms rather than
    custom algorithms (\cov{5.12}) both describe situations not easily detectable by our analysis, e.g.,
    our linter would have to figure out if a certain predicate indeed implements quicksort.
    \item[--] Augmenting Prolog with a run-time type and mode checking system as suggested in \cov{5.13} is a point
    we absolutely share. However, this again should not be part of a linter.
    Instead, we refer to the type systems suggested~\cite{10.1007/978-3-540-89982-259} .
\end{itemize}

The guidelines in \cov{6} focus on the general process of software development and testing and are thus less appropriate for source code analysis.
Two of them however can easily be checked automatically: avoiding constructs that are almost always wrong (\cov{6.5}) and detecting and avoiding magic constants (\cov{6.7}).

To summarize, the current selection of coding guidelines available for Prolog can only partially be verified and enforced automatically.
In particular, rules involving sense and meaning of comments or 
pronunciation cannot efficiently be verified.
To enforce the automatically verifiable rules, we implemented a linting tool which we will discuss 
in the following section.

\section{Static Source Code Analysis for Prolog}%
\label{sec:linter}

Tools for static source code analysis have been established in the 
process of software engineering in all major programming languages.
They allow developers to discover potential faults early, 
e.g., because misspelled variable names are never used, and highlight 
means for optimizations, e.g., to eliminate loop invariants.
In addition, static source code analysis is used to enforce the 
adherence to coding standards.


In contrast to imperative programming languages with a fixed 
set of keywords, Prolog's syntax is very flexible.
With self-defined operators, terms can be written without parentheses.
For instance, the term \mbox{\textit{``a b''}} in functional notation 
represents either the compound \textit{``a(b)''} or \textit{``b(a)''}, 
depending on whether \texttt{a/1} is a prefix operator or \texttt{b/1} 
is a postfix operator.
For code analysis this distinction is required, because some 
coding guidelines (e.g., \cov{3.5 and 3.6}) target predicates 
but not arguments.
Since the used operators 
could also be defined in imported modules, static analysis of Prolog programs often requires 
information collected at runtime.
Some Prolog systems, most notably Ciao Prolog, use refined module systems to allow for more thorough static analysis~\cite{Gras:2000:NMS:647482.725982}.


For Prolog systems in general, this information is available on compilation.
Therefore, a feasible approach to implement a linter for Prolog might 
be to use its built-in term expansion mechanism.
In the following, we discuss this technique as well as the traditional 
way to implement a linter using a parser.

\paragraph{Term Expansion}
Term expansion is a mechanism to rewrite Prolog code at compilation, similar to macros in other programming languages.
Although not part of the ISO Prolog standard, it is supported by most 
Prolog systems. When loading code into Prolog, its compiler calls 
\texttt{expand\_term/2} on each term read from the input.
The asserted term can be modified by defining the predicate 
\texttt{term\_expansion(+Term1,-Term2)} taking the 
original \texttt{Term1} and replaces it by \texttt{Term2}.
Both take account of operator definitions, i.e. functional notation is handled by the Prolog compiler.

Without modification of the term, \texttt{term\_expansion/2} 
provides a simple way to access all clauses on compilation.
Hence, it can be used to check adherence to the coding 
guidelines of \cov{3 and 5}, related to naming conventions 
and language idioms.
Layout (\cov{2}) cannot be 
checked this way, because only the Prolog terms are given to 
\texttt{term\_expansion/2}; all styling information is omitted, since 
it is not needed for compilation.%
\footnote{In \swipl and Quintus, the option 
\texttt{subterm\_positions} of \texttt{read\_term/2} provides only 
character positions which do not allow to correctly identify newlines.
On the other hand, SICStus' \texttt{layout} option provides only the 
line numbers of the subterms but not their actual indentations.
The original variable names can be accessed in all implementations, 
because the needed option \texttt{variable\_names} is part of the ISO 
Prolog standard.}
Source code annotations cannot be accessed in the term expansion.
Although \swipl allows to define the hook 
\texttt{prolog:comment\_hook/3}~\cite[Sec.~B.8]{swiref} to get access to
all comments, the tests regarding source code annotations (\cov{4})
would be split across this hook and the term expansions.

Another disadvantage of linting using term expansion is that 
\texttt{term\_expansion/2} is stateless.
Several coding conventions require knowledge about other clauses, which 
one would normally pass as an additional argument.
Instead, persistent information would have to be handled using 
dynamic predicates.

With term expansion, parsing is left 
to the compiler of the used Prolog system.%
\footnote{For automatic source code analysis -- for example in our 
empirical study in Section~\ref{sec:report} --, often source code from 
unknown sources is tested.
To prevent security issues when loading potential malicious code, we 
strongly recommend using the empty list \texttt{[]} for \texttt{Term2} 
in \texttt{term\_expansion/2}.
As a result, external code is neither asserted nor executed.}
This way, the linting rules can be defined by just processing Prolog 
terms, without having to implement a complete Prolog parser.

\paragraph{Parser}
Even though execution is not needed, term expansion can only be used to analyze code in valid Prolog syntax with respect 
to the used Prolog system.
Many~(all?) Prolog systems diverge from the ISO Prolog standard.%
\footnote{\textit{Conformity Testing I: Syntax}, list of ISO compliance 
for popular Prolog systems, collected by Ulrich Neumerkel: 
\url{http://www.complang.tuwien.ac.at/ulrich/iso-prolog/conformity_testing}%
.}
In consequence, support for multiple Prolog dialects and different Prolog systems cannot be 
achieved relying on term expansion, as it is restricted to the 
language features provided by the used system.
The same holds true for reusing existing parsers of different Prolog interpreters.
In contrast, with a dedicated Prolog parser, even source code designed for a 
different target system can be analyzed.

Although the parser loses runtime information available in the 
term expansion approach --~most importantly operator definitions~--, 
we opted for this choice. In the following, we present the linter's 
architecture, including a feature-rich Prolog parser.

\subsection{\textit{library(plammar)} -- A Prolog Parser and Serializer}
\label{sec:parser}

Static analysis of source code is typically 
split into two phases:
(i)~the lexical analysis, that converts a sequence of characters into a 
list of tokens,
and (ii)~the combination of tokens in order to generate a structural 
representation, often in form of a syntax tree.
In a third step, the concrete syntax tree~(CST) is converted into an 
abstract syntax tree~(AST), while removing and analyzing all layout 
information.

\paragraph{Lexical Analysis}
The syntax of Prolog is specified in the ISO Prolog 
standard~\cite{ISOProlog} using grammar rules in extended Backus--Naur 
form~(EBNF).
Nogatz et al. defined EBNF as an internal domain-specific language in 
Prolog~\cite{nogatz2019dcg4pt}.
Similar to DCGs, the grammar rules are translated into Prolog 
predicates with two additional arguments to hold lists of consumed 
resp. remaining symbols.
In contrast to the traditional term expansion scheme of DCGs, the 
resulting Prolog predicates also contain an additional argument to 
automatically store the corresponding parse tree, based on the rule's 
nonterminal.

\begin{figure}
	\begin{eqnarray*}
		\mathit{variable~token} & = & 
		\mathit{anonymous~variable}~|~\mathit{named~variable}~; \\
		\mathit{anonymous~variable} & = & 
		\mathit{variable~indicator~char}~; \\
		\mathit{named~variable} & = & 
		\mathit{variable~indicator~char},~\mathit{alphanumeric~char},~ 
		\{~\mathit{alphanumeric~char}~\} \\
		& | & \mathit{capital~letter~char}, 
		\{~\mathit{alphanumeric~char}~\}~; \\
		\mathit{variable~indicator~char} & = & 
		\mathit{underscore~char}~; \\
		\mathit{underscore~char} & = & "\_"~;
	\end{eqnarray*}
	\caption{Grammar rules for a variable token in Prolog as defined in 
	the ISO Prolog standard.}
	\label{lst:ebnf-example}
\end{figure}

As an example, the EBNF grammar rules for a variable token as defined 
in~\cite[Sec.~6.4.3]{ISOProlog} are given in 
Figure~\ref{lst:ebnf-example}.
It is translated into the predicate \texttt{variable\_token/3}, 
that creates a parse tree for a given input string and vice-versa.
In Figure~\ref{lst:variable-token-call}, we present an example query 
for the input string ``\texttt{\_a}''.%
\footnote{The ISO Prolog standard requires that a token shall not be 
followed by characters such that concatenating the characters of the 
token with these characters forms a valid token, i.e. only the second 
solution of Figure~\ref{lst:variable-token-call} is valid.
This requirement is realized in other parts of our parser.}
The syntax of other Prolog tokens is defined similarly.
As a result, our tool \textit{library(plammar)} provides the predicate 
\texttt{prolog\_tokens(?Source,?Tokens)} that takes Prolog source code 
and generates the list of tokens, and the other way around.

\begin{figure}
	\begin{Verbatim}[fontsize=\footnotesize]
?- variable_token(Parse_Tree, "_a", R).
% first solution, consuming only "_":
R = "a", Parse_Tree = variable_token(anonymous_variable(
                        variable_indicator_char(underscore_char('_'))));
% second solution, consuming the whole string "_a":
R = "",  Parse_Tree = variable_token(named_variable([
                        variable_indicator_char(underscore_char('_')),
                        alphanumeric_char(alpha_char(letter_char(
                          small_letter_char(a) ))) ])) .
\end{Verbatim}
\caption{Usage example for the generated \texttt{variable\_token/3} 
with the input string \texttt{"\_a"}.}
\label{lst:variable-token-call}
\end{figure}

\paragraph{Parsing}

The ISO Prolog standard specifies how to combine tokens into valid 
terms with EBNF again.
Therefore, the list of tokens generated by \texttt{prolog\_tokens/3} 
can be processed as before, consuming tokens rather than characters.
The generated parse tree strictly follows the structure of the applied 
grammar rules, with its nonterminals as subtrees.
\cite{nogatz2019dcg4pt} provide a more detailed discussion of aspects 
when parsing Prolog code this way, e.g., to correctly handle operator 
precedences.

\paragraph{Towards an Abstract Syntax Tree}

The generic structure of the parse tree is based on the 
grammar's nonterminals, which allows to easily follow the application 
of the EBNF rules in the ISO Prolog standard.
On the other hand, this concrete syntax tree (CST) is very verbose.
The contained layout information are only required to test the 
adherence to \cov{2}; all other coding guidelines are applicable to an 
abstract syntax tree (AST).
We therefore define a Prolog predicate that transforms a parse tree 
into its corresponding AST, and the other way around.
For instance, a program with just a single rule returns the following 
AST:
\begin{center}
\begin{Verbatim}[fontsize=\footnotesize]
  ?- prolog_ast(string("positive(X) :- X > 0."), AST).
  AST = prolog([ rule(                                            % single rule
                   compound(atom(positive), [variable('X')]),            % head
                   [infix(>, xfx, variable('X'), integer(0))] )]).       % body
\end{Verbatim}
\end{center}

\subsection{Rules for Code Formatting and Code Quality}
\label{sec:linting}

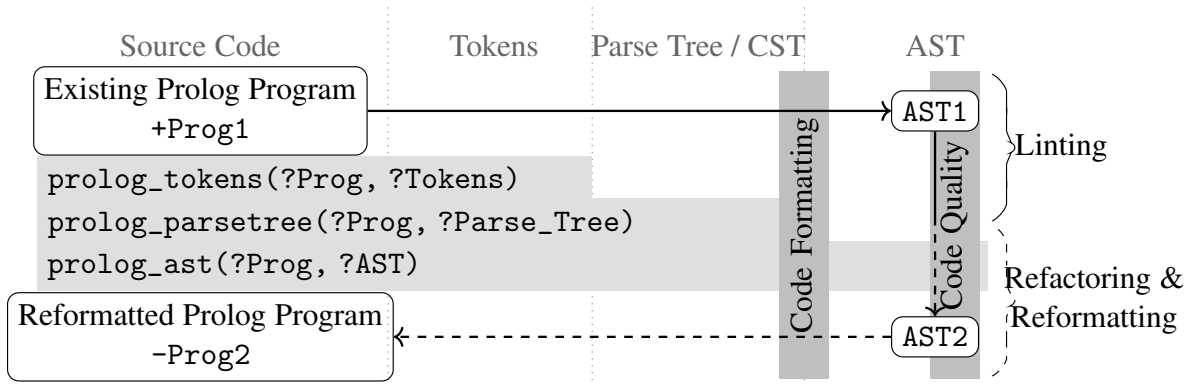
\begin{figure}[t]
	\resizebox{\linewidth}{!}{
		\begin{tikzpicture}
		[rounded corners]
		
		\node [opacity=0.6] at (1.5,4.3) {Source Code};
		\draw [dotted, opacity=0.5] (3.8,0.2) -- (3.8,4.8);
		
		\node [opacity=0.6] at (5.1,4.3) {Tokens};
		\draw [dotted, opacity=0.5] (6.3,0.2) -- (6.3,4.8);
		
		\node [opacity=0.6] at (7.6,4.3) {Parse Tree / CST};
		\draw [dotted, opacity=0.5] (8.9,0.2) -- (8.9,4.8);
		
		\node [opacity=0.6] at (10.5,4.3) {AST};
		
		\node [fill=gray!25,rounded corners=0,text 
		width=186,align=left] at (2.90,2.64) 
		{\texttt{prolog\_tokens(?Prog, ?Tokens)}};
		
		\node [fill=gray!25,rounded corners=0,text 
		width=260,align=left] at (4.2,2.12) 
		{\texttt{prolog\_parsetree(?Prog, ?Parse\_Tree)}};
		
		\node [fill=gray!25,rounded corners=0,text 
		width=324,align=left] at (5.32,1.6) 
		{\texttt{prolog\_ast(?Prog, ?AST)}};	
		
		\node[draw, rectangle, align=center] (source1) at (1.5,3.5) 
		{Existing Prolog Program \\ \texttt{+Prog1}};
		\node[draw, rectangle, align=center] (source2) at (1.5,0.73) 
		{Reformatted Prolog Program \\ \texttt{-Prog2}};
		
		\node [rotate=90,fill=gray!50,rounded 
		corners=0,text width=100,align=center] at (8.9,2.1) 
		{Code Formatting};	
		\node [rotate=90,fill=gray!50,rounded 
		corners=0,text width=100,align=center] at (10.74,2.1) 
		{Code Quality};	
	
		\node[draw, rectangle, fill=white] (ast1) at (10.5,3.5) 
		{\texttt{AST1}};
		\node[draw, rectangle, fill=white] (ast2) at (10.5,0.73) 
		{\texttt{AST2}};
		
		\draw [->, thick] (source1) to (ast1);
		\draw [-, thick] (ast1) to (10.5,2.2);
		\draw [->, thick, dashed] (10.5,2.2) to (ast2);
		\draw [->, thick, dashed] (ast2) to (source2);
		
		\draw [decorate,decoration={brace,amplitude=10pt,mirror}] 
		(11.2,2.15) -- (11.2,3.95) node [black,midway] {};
		\node [] at (12.05,3.05) {Linting};
		\draw 
		[decorate,dashed,decoration={brace,amplitude=10pt,mirror}] 
		(11.2,0.25) -- (11.2,2.10) node [black,midway] {};
		\node [align=center] at (12.45,1.18) {Refactoring \& \\ 
		Reformatting};
		\end{tikzpicture}
	}
	\caption{Architecture Overview for Linting with 
	\textit{library(plammar)}.}\label{fig:architecture}
\end{figure}

Figure~\ref{fig:architecture} summarizes the parser's overall 
architecture.
For each step, \textit{library(plammar)} provides a predicate:
(i)~\texttt{prolog\_tokens/2} holds the tokens for the given Prolog source 
code; \linebreak
(ii)~\texttt{prolog\_parsetree/2} generates the corresponding parse 
tree;
and (iii)~\texttt{prolog\_ast/2} additionally transforms the CST into 
AST.
Every predicate can take an additional argument to pass options, 
e.g., to specify the targeted Prolog system or dialect.

In general, we distinguish between \textit{code formatting rules} and 
\textit{code quality rules}.
The first are related only to the CST; changes are not represented in 
the program's AST, code 
formatting rules are checked on-the-fly in the tree transformation step 
of \texttt{prolog\_ast/2}.

%
Most Prolog tokens are allowed to be preceded by a 
\textit{layout text sequence}~\cite[Sec.~6.4.1]{ISOProlog}, i.e., a 
sequence of whitespaces, newlines, or comments.
In the transformation step from CST to AST, this layout information is 
removed but tested to its adherence to the given guidelines.
The preferred formatting can be specified as a list of 
options, also 
referred to by the coding guidelines by Covington et al.:
\begin{itemize}
	\item[--] \cov{2.1} and \cov{2.2}: \textit{indent(Integer)},
	\item[--] \cov{2.3}: \textit{max\_line\_length(Integer)},
	\item[--] \cov{2.4}: \textit{max\_subgoals(Integer)} and 
	\textit{max\_rule\_lines(Integer)},
	\item[--] \cov{2.5}: \textit{space\_after\_arglist\_comma(yes/no)},
	\item[--] \cov{2.6}: \textit{newline\_after\_clause(yes/no)},
	\item[--] \cov{2.7}: \textit{newline\_after\_rule\_op(yes/no)} and
				\textit{newline\_after\_subgoal(yes/no)},
	\item[--] \cov{2.14}: 
	\textit{indent\_between\_repeat\_cut(yes/no)}
\end{itemize}

Additionally, we provide options for common best-practices, e.g., 
to prevent trailing whitespaces.


In contrast to formatting rules, code quality rules consider only the AST.
This distinction is also reflected in the implementation: code quality rules are defined separately as 
a tree traversal for the AST.
Currently, they are simply checks, without further transformations.
We support options to handle \cov{3.1}, \cov{3.4} and \cov{3.12}.

\cov{4} suggest style guidelines for source code comments, \cov{5} and 
\cov{6} recommend the usage and avoidance of several language idioms.
These are not yet common in the Prolog community, though we want 
to add checks for them in the future.

\subsection{Benefits from Prolog Features}
\label{sec:prolog-features}

Our implementation shows that Prolog is particularly suited for 
implementing linters and software analysis tools in general.
In particular, the following features came in handy for us:

\begin{itemize}
	\item[--] \textit{Definite clause grammars.}
	The definition of Prolog's syntax with EBNF rules resembles DCGs.
	As a result, the grammars for tokens and terms can be directly 
	adopted from the ISO Prolog standard, resulting in an executable 
	Prolog program after minor adjustments.
	
	\item[--] \textit{Backtracking.}
	For performance reasons, the parser avoids backtracking where 
	possible; all grammar rules require at most a small look-ahead.
  Only in cases where the program's environment is underdetermined,
  e.g., if the targeted Prolog dialect is unknown, or operator 
	definitions are missing, backtracking can be used.
	This way, the program ``\textit{a b.}'' can be parsed, deducing 
	that \texttt{a/1} has to be defined as prefix operator, or 
	\texttt{b/1} as postfix operator.%
	\footnote{In our empirical research of Section~\ref{sec:report}, we 
	deactivated this option to deduce missing operator definitions, as 
	it slows down the parsing process because of the bigger search 
	space.}
	
	\item[--] \textit{Logic Variables.}
	The settings we presented in Section~\ref{sec:linting} also support 
	logic variables as arguments.
	As a result, it is bound to the correct value where possible.
	For instance, a variable in \textit{max\_line\_length} is bound to 
	the maximum value of all observed line lengths.
	
	\item[--] \textit{Predicates as relations.}
	Each predicate of \textit{library(plammar)} expresses a relation 
	between Prolog source code and its corresponding list of tokens, 
	parse tree, or AST.
	For this purpose, the term expansion creating Prolog predicates 
	based on EBNF was designed to use only pure predicates.
	\textit{library(plammar)} can therefore be used in both directions 
	without modification.
\end{itemize}

Currently, violations of coding guidelines are returned as a 
list of warnings or just printed when calling \texttt{prolog\_ast/3} 
(code formatting rules) and \texttt{check\_ast/2} (code quality 
rules):
\begin{center}
	\begin{Verbatim}[fontsize=\small]
  ?- prolog_ast(file('/tmp/in.pl'), AST, Options), check_ast(AST, Options).
	\end{Verbatim}
\end{center}
\noindent

The fact that \textit{library(plammar)} can be used in reverse simplifies the extension to repair of rule violations.
In this case, one just has to provide \texttt{Parse\_Options} with 
non-strict settings for parsing, and \texttt{Serialize\_Options} with 
the preferred style settings for serializing. Instead of 
\texttt{check\_ast/2}, a predicate \texttt{transform\_ast/3} has to be 
used that transforms the program's AST according to the coding 
guidelines:
\begin{center}
	\begin{Verbatim}[fontsize=\small]
  ?- prolog_ast(file('/tmp/in.pl'), AST_In, Parse_Options),
     transform_ast(AST_In, AST_Out, Serialize_Options),
     prolog_ast(file('/tmp/out.pl'), AST_Out, Serialize_Options).
	\end{Verbatim}
\end{center}

\subsection{Support for \swipl Extensions}
\label{sec:swi-syntax}

We opted for a dedicated Prolog parser to support different Prolog 
systems and dialects.
While we wanted to support as many Prolog systems as possible,
we were particularly interested in \swipl.
With the release of version 7, \swipl further 
diverged from the ISO Prolog standard~\cite{wielemaker2014swi}.
We implemented 20 options to enable or disable non-ISO-conform language 
features supported by recent versions of \swipl.
Among others, \textit{library(plammar)} considers the following 
settings, each as \textit{yes/no}:

\begin{itemize}
	\item[--] \textit{dicts}:
	This adds a new token for structures with 
	named arguments, e.g., ``\texttt{point\{\,a:\,1\,\}}''.
	In addition, the operator ``\texttt{.}'' is added for field 
	extraction.
	
	\item[--] \textit{allow\_compounds\_with\_zero\_arguments}:
	Allow compounds without arguments, e.g., ``\texttt{a()}''.
	
	\item[--] \textit{allow\_arg\_precedence\_geq\_1000} and 
	\textit{allow\_operator\_as\_operand}:
	\swipl allows for terms that would require parenthesis 
	according to the ISO Prolog standard, e.g., ``\texttt{X = -}'', and 
	\linebreak\mbox{``\texttt{[a :- b]}''} instead of ``\texttt{[(a :- b)]}''.
	
	\item[--] \textit{allow\_integer\_exponential\_notation}:
	The ISO Prolog standard requires one to write ``\texttt{1.0e3}'' 
	instead of ``\texttt{1e3}''.
\end{itemize}

The same approach could be followed to adapt to further Prolog systems and dialects.
In our static code analysis, we only consider the program's syntax.
Although possible, we do not yet check for system-specific built-in 
predicates.

\section{Status Report}
\label{sec:report}

Even though there were several pleas for the consistent usage of coding guidelines in the Prolog community,
it remains unclear, whether there is consensus on how to handle formatting
and specific coding constructs.
Several possible scenarios come to mind:

\begin{enumerate}
\item Most of the community adheres to one batch of guidelines, e.g.,
    to Covington et al.~(2012).
\item The community is fractured, adhering to different guidelines.
    One partitioning might be, that groups using \swipl
    write code differently than groups using SICStus.
    Another possibility might be that certain projects, e.g., packages 
    shipped with \swipl,
    enforce a set of rules concerning the code.
\item
    Programmers have an individual, consistent coding style that they stick to.
\item
    Individual programmers do not care and mix different coding styles in their own projects.
\end{enumerate}

We suspect that contributors working on larger, serious projects
        mostly adhere to a project-wide style,
        though there are some minor deviations to be expected.
In contrast, individuals managing a small code basis, e.g., homework repositories,
        usually are not concerned with code style,
        mostly because they have not considered any guidelines
        or the project is not serious enough.

In the following, we use our tool
to determine whether the Prolog community
adheres to any guidelines
and which of our hypotheses
will be falsified when checked against reality.

\subsection{Empirical Research}

For our analysis,
we consider 
the \swipl \enquote{batteries-included} 
packages\footnote{\url{https://github.com/SWI-Prolog/swipl-devel/tree/9afb9384bb/packages}}
as well as
the list of known \swipl
packages\footnote{\url{http://www.swi-prolog.org/pack/list}}.
A similar analysis for other dialects such as SICStus or Ciao is possible using the same grammar and linter as discussed above and will part of our future research.
We discard unavailable packages and files only consisting of facts.

For all files,
we set a 10-second timeout.
We excluded source files larger than 1 MB or longer than 20.000 lines of code,
as they are usually generated by some tool.
Additionally, we excluded some packages due to
inconsistent operator definitions within the package,
invalid Prolog code, and
operator definitions that are not supported any longer
by \swipl.%
\footnote{A list of tested packages is available at 
\url{https://github.com/fnogatz/plammar-community-evaluation}.}

\paragraph{Community Packages}
The considered code base consists of 3815 Prolog files from 251 
packages, of which 2775 contained rules.
More than 90 \% of the files could be parsed by our tool.
We elaborate on why others could not~\cref{sec:limitations}.

\paragraph{SWI Packages}
Packages shipped with \swipl consist of 688 Prolog files,
609 of which could be processed
and 530 containing at least one rule.
They stem from 34 packages
and consist of 137051 lines of code.



\begin{figure}[t]
    \centering
\begin{subfigure}{.49\textwidth}
    \resizebox{\textwidth}{!}{
\begin{tikzpicture}
\begin{axis}[
    axis y line*=left,
    ylabel=Number of Files (Community Packages),
    xlabel=Maximal Number of Subgoals,
    ybar,
    ymin=0,ymax=554,
    xmin=-0.5, xmax=25.5,
    extra x ticks=25, extra x tick labels={25+}, extra x tick style={tick label style={fill=white}},
]
\addplot +[
    fill=red!60,opacity=0.8,
    draw=red,
    pattern=north east lines,
    pattern color = red,
    hist={
        bins=26,
        data min=-0.5,
        data max=25.5,
    }
    ] table [y index=0] {subgoals.data};
    \label{plot-community}
\addlegendentry{Community}
\end{axis}
\begin{axis}[
    axis x line=none,
    axis y line*=right,
    ylabel near ticks, yticklabel pos=right,
    ylabel=Number of Files (SWI Packages),
    ybar,
    ymin=0,ymax=100,
    xmin=-0.5, xmax=25.5
]
\addlegendimage{/pgfplots/refstyle=plot-community}\addlegendentry{Community}
\addplot +[
    fill=blue!60,opacity=0.4,
    hist={
        bins=26,
        data min=-0.5,
        data max=25.5,
    }
] table [y index=0] {subgoals-swi.data};
    \label{plot-swi}
\addlegendentry{SWI}
\end{axis}

\end{tikzpicture}
}
\caption{Maximal Number of Subgoals}%
\label{fig:subgoal}
\end{subfigure}
\begin{subfigure}{.49\textwidth}
    \resizebox{\textwidth}{!}{
\begin{tikzpicture}
\begin{axis}[
    axis y line*=left,
    ylabel=Number of Files (Community Packages),
    xlabel=Maximal Lines per Subgoal,
    ybar,
    ymin=0,ymax=554,
    xmin=-0.5, xmax=25.5,
    extra x ticks=25, extra x tick labels={25+}, extra x tick style={tick label style={fill=white}},
]
\addplot +[
    draw=red,
    pattern=north east lines,
    pattern color = red,
    hist={
        bins=26,
        data min=-0.5,
        data max=25.5,
    }
    ] table [y index=0] {maxrules.data};
    \label{plot-community2}
\addlegendentry{Community}
\end{axis}
\begin{axis}[
    axis x line=none,
    axis y line*=right,
    ylabel near ticks, yticklabel pos=right,
    ylabel=Number of Files (SWI Packages),
    ybar,
    ymin=0,ymax=100,
    xmin=-0.5, xmax=25.5
]
\addlegendimage{/pgfplots/refstyle=plot-community2}\addlegendentry{Community}
\addplot +[
    fill=blue!60,opacity=0.5,
    hist={
        bins=26,
        data min=-0.5,
        data max=25.5,
    }
] table [y index=0] {maxrules-swi.data};
    \label{plot-swi2}
\addlegendentry{SWI}
\end{axis}

\end{tikzpicture}
}
\caption{Maximal Number of Lines per Rule}%
\label{fig:linenums}
\end{subfigure}

    \begin{subfigure}{.49\textwidth}
\def\angle{0}
\def\radius{2}
\def\cyclelist{{"orange","blue","red","green"}}
\newcount\cyclecount \cyclecount=-1
\newcount\ind \ind=-1
        \resizebox{\textwidth}{!}{
\begin{tikzpicture}[nodes = {font=\sffamily}]
  \foreach \percent/\name in {
      28.3/Mixed,
      43.5/Spaces Only,
      23.8/Tabs Only,
      4.4/No Indentation
    } {
      \ifx\percent\empty\else               
        \global\advance\cyclecount by 1     
        \global\advance\ind by 1            
        \ifnum3<\cyclecount                 
          \global\cyclecount=0              
          \global\ind=0                     
        \fi
        \pgfmathparse{\cyclelist[\the\ind]} 
        \edef\color{\pgfmathresult}         
        \draw[fill={\color!50},draw={\color}] (0,0) -- (\angle:\radius)
          arc (\angle:\angle+\percent*3.6:\radius) -- cycle;
        \node at (\angle+0.5*\percent*3.6:0.7*\radius) {\percent\,\%};
        \node[pin=\angle+0.5*\percent*3.6:\name]
          at (\angle+0.5*\percent*3.6:\radius) {};
        \pgfmathparse{\angle+\percent*3.6}  
        \xdef\angle{\pgfmathresult}         
      \fi
    };
\end{tikzpicture}
        }
\caption{Indentation (Community Packages)}%
\label{fig:indent-community}
\end{subfigure}
\begin{subfigure}{.49\textwidth}
\def\angle{0}
\def\radius{2}
\def\cyclelist{{"orange","blue","red"}}
\newcount\cyclecount \cyclecount=-1
\newcount\ind \ind=-1
        \resizebox{\textwidth}{!}{
\begin{tikzpicture}[nodes = {font=\sffamily}]
  \foreach \percent/\name in {
      9.1/Mixed,
      89.4/Spaces Only,
      1.5/Tabs Only,
    } {
      \ifx\percent\empty\else               
        \global\advance\cyclecount by 1     
        \global\advance\ind by 1            
        \ifnum3<\cyclecount                 
          \global\cyclecount=0              
          \global\ind=0                     
        \fi
        \pgfmathparse{\cyclelist[\the\ind]} 
        \edef\color{\pgfmathresult}         
        \draw[fill={\color!50},draw={\color}] (0,0) -- (\angle:\radius)
          arc (\angle:\angle+\percent*3.6:\radius) -- cycle;
        \node at (\angle+0.5*\percent*3.6:0.7*\radius) {\percent\,\%};
        \node[pin=\angle+0.5*\percent*3.6:\name]
          at (\angle+0.5*\percent*3.6:\radius) {};
        \pgfmathparse{\angle+\percent*3.6}  
        \xdef\angle{\pgfmathresult}         
      \fi
    };
\end{tikzpicture}
        }
\caption{Indentation (SWI batteries included)}%
\label{fig:indent-swi}
\end{subfigure}
\caption{Properties per File.}
\end{figure}
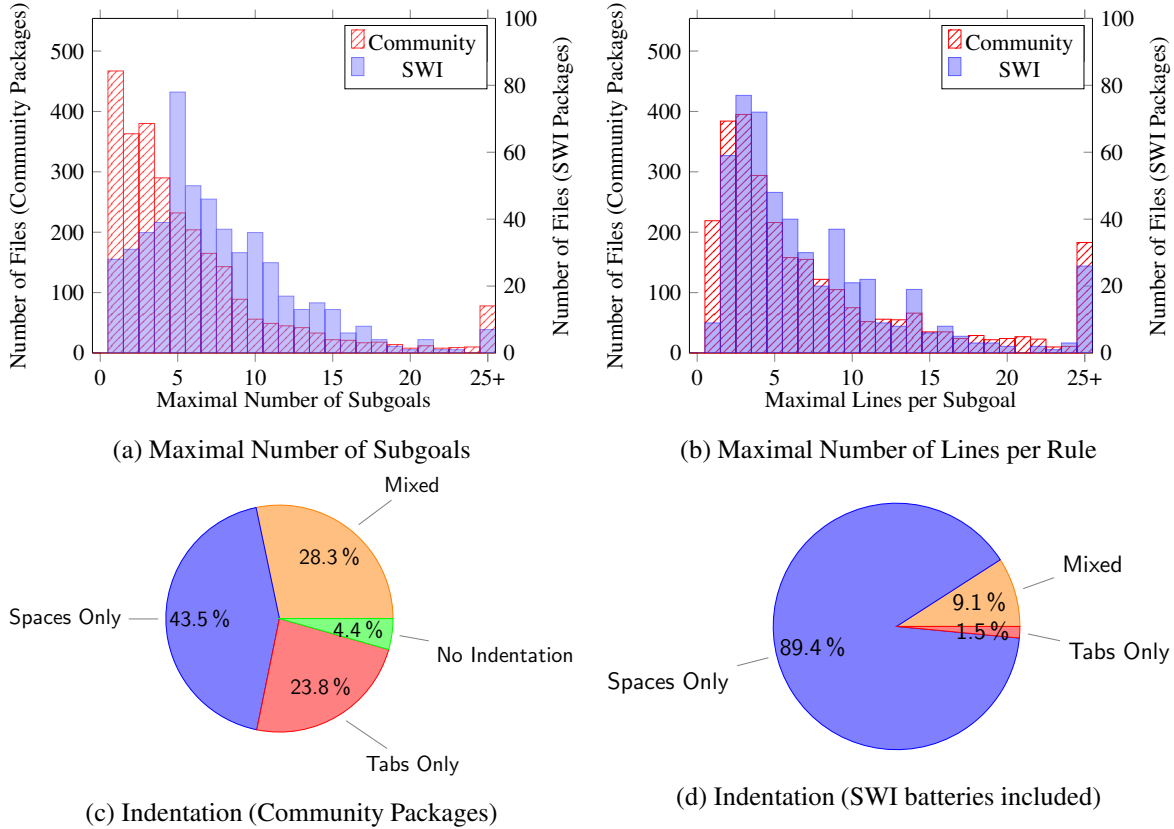

Following,
we present the results for some coding guidelines
(\cov{2.1, 2.2}) in detail.
Among the most discussed is how to properly indent:
\cref{fig:indent-community,fig:indent-swi} show the share of different 
styles,
i.e., no indentation vs.
indentation with spaces or tabs
that is consistent in the sense that spaces and tabs are not mixed,
but not that the indentation size is the same.
Lastly, both tabs and spaces might be used for indentation.
As can be seen, there is no clear trend.

Limiting the horizontal length of source code lines 
is considered a best practice in~\cov{2.3}.
From 361782 source code lines overall (including blank lines and comments),
only 12431 lines (3.4 \%) are longer than 80 characters.
However, violations of this rule can be found in 879 files (31.6 \%).
In the \enquote{batteries included} packages,
this figure is slightly smaller,
with 756 out of 137051 lines (0.6 \%) in
92 of the 530 (14.0 \%) analyzed files
exceeding 80 characters.

The length of clauses should be limited as well,
both in lines and sub-goals.
\cref{fig:subgoal} and \cref{fig:linenums}
show the distribution of the largest clause per file.
Most clauses are rather short,
as many are facts or dispatch to another predicate.
There are a few outliers ranging over more than
a hundred lines and sub-goals,
though most Prolog programmers seem to adhere to a sane limitation.

There are also many whitespace rules suggested by \textsc{Cov},
including usage of space(s) after commas in argument lists of a goal (cf. \cov{2.5}).
In the community packages,
a large fraction rejects this idea,
resulting in 168626 missing spacings on 361782 (46.6 \%) lines of code.
In the \enquote{batteries included} packages,
there are 24762 instances on 137051 (18.1 \%) of missing whitespace.
This suggests that there might be a coding guideline
for the \swipl packages.

We can also find minor differences in superfluous whitespaces
at the end of a line (2.4 \% vs. 1.1 \%),
missing linebreaks after the rule operator ``\verb|:-|''
(6.0 \% vs. 1.0 \%),
and missing newlines after a sub-goal (6.9 \% vs. 0.4\%)
when comparing the community (first number)
and the \enquote{batteries included} (second number) packages.
However, there are a few instances
of a missing newline after a clause
in the community packages (0.1 \%),
while there are none in \enquote{batteries included}.

\subsection{SWI-Specific Features}

\swipl includes many syntactic elements
not part of ISO Prolog.
Thus, we are interested
in how far they have been adopted.
In \cref{tbl:features}, the amount of usages of several features
is given, again partitioned into community packages
and \enquote{batteries included} packages.

\begin{table}
\caption{Adaptation of SWI-Specific Features in the Prolog Community.}
\label{tbl:features}
\resizebox{\textwidth}{!}{
\begin{tabular}{llrrrr}
    \toprule
    ~ & ~ & \multicolumn{2}{c}{Community Packages} & 
    \multicolumn{2}{c}{Batteries Included}       \\
    \cmidrule{3-4} \cmidrule{5-6}
    Feature & Example & Occurr. & in \textunderscore{} of 251 packs 
    & Occurr. & in \textunderscore{} of 34 packs \\ 
    \midrule
    shebang & \exm{\#!swipl} & 11 & 6 (2.4 \%) & 2 & 
    2 (5.9 
    \%) \\
    digit groups & \exm{1\_000} & 77  & 11 (4.4 \%)& 4 & 2 (5.9 \%)\\
    dicts & \exm{a\{b:1\}} & 883 & 35 (13.9 \%)& 119 & 10 (24.4 \%)\\
    unicode character escape & \exm{\textbackslash u2C6F} & 13 & 2 (0.8 
    \%)& 50 & 7 
    (20.6 \%)\\
    missing closing ``\textbackslash'' & \exm{\textbackslash x00} & 
    9 & 3 
    (1.2 
    \%)& 0 & 0 (0.0 \%)\\
    single quote char constant & \exm{0''} & 13 & 5 (2.0 \%)& 1 & 1 
    (2.9 \%)\\
    zero arguments compound & \exm{pi()} & 60 & 8 (3.2 \%)& 0 & 0 (0.0 
    \%)\\
    tab in quotes & \exm{"\textbackslash t"} & 159 & 9 (3.6 \%)& 0 & 0 
    (0.0 \%)\\
    integer exp. notation & \exm{1e3} & 38 & 8 (3.2 \%)& 16 & 3 (8.8 
    \%)\\
 \bottomrule
\end{tabular}
}
\end{table}

As can be seen, most features are not very widespread yet.
One reason is that some syntactic elements are very situational,
e.g., usage of tabulators in quotes,
whereas others have suitable alternatives,
e.g., the additional exponential notation.
Additionally, these features were introduced fairly recent
while many packages already exist significantly longer.
Thus, it is quite unlikely that existing code is re-written
in order to use new language features.

Yet, dicts in \swipl seem to be an exception:
around 14 \% of community packages and 25 \% of those shipped with 
\swipl make use of dicts at some point.
This is an indicator of both active development
and interest in the community to use proper associative data structures.

%

%
%



\subsection{Limitations of Analysis}
\label{sec:limitations}

Although to the best of our knowledge, our tool 
\textit{library(plammar)} is compliant to the ISO Prolog standard, it 
has the following known limitations:

\begin{itemize}
	\item[--] \swipl and others offers full Unicode support.
	Although we handle character escapes like
	\verb@\u2C6F@ for ``$\forall$'', we do not yet support the
	literal appearance of ``$\forall$'' as part of a comment, atom, 
	etc.
	
	\item[--] \swipl allows ``[]'' and ``\{\}'' to be defined as block 
	operators~\cite[Sec.~5.3.3]{swiref}. In addition, \textit{quasi 
	quotations} have been added in 2003~\cite{wielemaker2013s}.
	Both techniques are often used for embedding domain-specific 
	languages into Prolog, but are not yet supported by 
	\textit{library(plammar)}.
	
	\item[--] In contrast to the ISO standard, \swipl supports nesting bracketed comments.
\end{itemize}

We are working to remove these limitations in order to
reduce the current rate of 10~\% of the files that cannot be processed. 
Besides the missing support for the stated language extensions, there 
are some general problems that occur on static source code analysis for 
Prolog:

\begin{itemize}
	\item[--] \textit{No explicit export of operator definitions.}
	In general, parsing Prolog 
	requires knowing all operator definitions within a 
	package.
	Although Part II of the ISO standard on Prolog's module 
	system introduces the directive \texttt{module/2}, used 
	to declare a module and all its exported operator definitions,
	a module's entry points remain unclear.
	Hence, for our empirical research, we collected all operator 
	definitions given in the package's files first. 
	Otherwise, one had to analyze the package's call hierarchy.
	This is impossible in general, as the module system allows accessing 
  (possibly) internal module files.
	
	\item[--] \textit{No explicit import of operator definitions.}
	In addition, some of \swipl's ``batteries included'' and community 
	packages introduce new operators.
	For our empirical research, we therefore created a list of popular 
	packages with their exported operators.%
	
	\item[--] \textit{No explicit engines.}
	In our empirical research, we treated all Prolog files as if 
	designed for \swipl. This is no restriction, 
	because the language extensions introduced in 
	Section~\ref{sec:swi-syntax} are backward compatible to the ISO 
	Prolog standard.
	Nevertheless, it is desirable to make the targeted Prolog system 
	explicit to allow testing for specific 
	language features.
	For instance, digit groups have been added to \swipl of version~7, 
	whereas the definition of the back quote ``\texttt{\`{}}'' as an 
	operator was only allowed up to version~6.
\end{itemize}

These problems could be addressed by a more powerful package system, as 
suggested in~\cite{wielemaker152swi} or~\cite{Gras:2000:NMS:647482.725982}.
It would help to resolve dependencies including their operator 
definitions.
A simple solution could be to add suchlike properties to a module 
meta-description.

\section{Related Work}
\label{sec:related-work}

The importance of coding guidelines has been widely accepted throughout different programming languages and paradigms.
Existing collections of guidelines range from academic papers, such as the ones we mentioned, to whole books, e.g., in case of Java~\cite{long2013java}.
Even for low-level languages such as specialized assembly codes, coding guidelines have been suggested~\cite{osti6232072}.
In particular, several coding guideline have been suggested with security in mind, e.g., for C~\cite{Seacord:2008:CCS:1502209,guidelinesforsecurecoding}.

Quite often, adherence to coding guidelines is improved by relying on tools.
For Prolog, there is only a small number of existing 
development tools for linting as well as code reformatting.
Logtalk~\cite{de2003logtalk}, an object-oriented logic programming 
language that extends and leverages Prolog, ships with several compiler-based checks that can be also applied to standard Prolog.
Integrations for Prolog into popular IDEs such as Visual Studio Code%
\footnote{VSC-Prolog: \url{https://github.com/arthwang/vsc-prolog}}
make use of \swipl's introspection capabilities, and SWI-specific 
predicates for code formatting, e.g., \texttt{portray\_clause/2}.
Prolog implementations of the language server protocol~(LSP)%
\footnote{LSP: 
\url{https://microsoft.github.io/language-server-protocol/}},
which provides a unified API that is supported by most IDEs, follow the 
same approach%
\footnote{E.g., \url{https://github.com/jamesnvc/lsp_server},
and \url{https://github.com/LukasLeppich/prolog-vim}}.

Parsing of Prolog by expanding the EBNF rules given in the ISO Prolog 
standard by an additional parse tree argument has been 
suggested by
Nogatz et al. (2019).
Seipel et al. (2003)
analyze and visualize Prolog source code 
whose parse tree is given in XML. This representation is the base for 
program refactorings~\cite{hopfner2005aprolog}.
A more detailed catalogue of Prolog refactorings is presented 
in~\cite{serebrenik2008improving}.

Analyzing Prolog code by only using term expansions as introduced in
\cref{sec:linter} has been the base for semi-automatic refactoring
of large Prolog systems.
In~\cite{mera2013porting}, an expert system consisting of almost
1~million lines of Prolog code was ported to \swipl after specifying
appropriate rewrite rules for compatibility.
However, the changes are applied just at compile-time, without any
modifications on the original source code.
With \textit{library(plammar)}, existing code can be reformatted
according to the coding guidelines.

\section{Conclusion and Future Work}
\label{sec:conclusion}

In summary, we presented a linter for Prolog, that 
supports the coding guidelines presented by Covington et al. (2012)
as far as we deemed feasible.
While certain rules could easily be checked automatically, other remain 
out of reach for now.
Our tool \textit{library(plammar)} is published as \swipl package
and on GitHub at \url{https://github.com/fnogatz/plammar} (MIT License).

In consequence, we see three major directions for future research.
First, we want to broaden the scope of language features our linter can 
be applied to.
In particular, we want to add support for linting Constraint Handling 
Rules (CHR)~\cite{fruhwirth1998theory}
and \textit{library(clpfd)}~\cite{Triska12}.
To do so, we have to carefully consider, how our linter interacts with Prolog's term expansion capabilities.
Furthermore, the current set of coding guidelines should be extended to 
take into account CHR and CLP(FD) idioms.
For instance, to the best of our knowledge, there is not yet a 
developer's standard on how to format non-trivial rules in CHR.

Second, we intent to extend the number of rules we can check 
automatically.
As outlined above, language processing technologies might enable 
checking rules regarding pronunciation.
Other technique from NLP or argumentation analysis might help to partially capture the quality of certain comments~\cite{10.1007/978-3-642-13881-27}.
Both extensions would greatly improve the user experience and should lead to Prolog code that is easier to understand and debug.
So far, we do not know about other linters implementing such 
techniques.

Finally, coding guidelines are best when they are easy to implement.
Therefore, we intent to improve developer tools for integrating 
\textit{library(plammar)}, e.g., to add support for LSP, and ease its 
usage in a CI pipeline.

\nocite{McLaughlin:1990:PRP:101356.101360}
\nocite{Kaplan:1991:PRP:122179.122184}

\bibliographystyle{eptcs}
\bibliography{article}

\label{lastpage}
\end{document}